\documentclass[twocolumn,showpacs,preprintnumbers,amsmath,amssymb,superscriptaddress]{revtex4}

\usepackage{epsfig}
\usepackage{graphicx}
\usepackage{dcolumn}
\usepackage{bm} 
\newcommand{\beq}{\begin{equation}}
\newcommand{\eeq}{\end{equation}}
\newcommand{\beqa}{\begin{eqnarray}}
\newcommand{\eeqa}{\end{eqnarray}}

\usepackage{subfig}

\begin{document}

\title{Rydberg states and momentum bifurcation in tunnel ionization with elliptically polarized light}

\author{A. S. Landsman}
\author{A. N. Pfeiffer}
\author{M. Smolarski}
\author{C. Cirelli}
\author{U. Keller}
\affiliation{ Physics Department, ETH Zurich, CH-8093 Zurich, Switzerland }
\email[corresponding author:  ]{alexandra.landsman@phys.ethz.ch}

\date{\today}

\begin{abstract} 
It is well-known from numerical and experimental results that the fraction of Rydberg states (excited neutral atoms) created by tunnel ionization declines dramatically with increasing ellipticity of laser light.   We present a method to analyze this dependence on ellipticity, deriving a probability distribution of Rydberg states that agrees closely with recent experimental \cite{b1} and numerical results.  In particular, contradicting the existing proposed mechanism \cite{b1} and general expectations, our results indicate that the dependence of Rydberg yield on laser ellipticity is not caused by a rescattering process.  Rather, we show that most Rydberg electrons never come back to the vicinity of the exit point after ionization, and end up relatively far from the atom (compared to the exit point) after the laser pulse has passed. 
We also present experimental data revealing a bifurcation that corresponds to a cut-off in Rydberg generation, and present an analytic derivation with a perturbative inclusion of the Coloumb force.
\end{abstract}


\maketitle


Tunnel ionization \cite{b3}, which can occur when an ionizing laser field strength is comparable to the Coulomb force binding the outer electron to its atom, is behind many recent break-throughs in attosecond science.  The 3-step model, which involves tunnel ionization, classical propagation and recollision \cite{b2}, works surprisingly well in clarifying many phenomena, most notably High Harmonic Generation (HHG) \cite{b34,b35} and Non-Sequential Multiple Ionization (NSMI) \cite{b5,b36,b37} .  Recently, the role of Coulomb force in laser ionization has drawn considerable attention from the strong field atomic physics community.  In particular, it is believed that to properly study many phenomena (e.g. focusing \cite{b23}, low energy structures at mid-infrared wavelengths \cite{b25}, certain asymmetries in momentum spectra \cite{b28}, tunneling geometry \cite{b29} and most notably for our analysis, the generation of Rydberg states \cite{b1,b21}), the inclusion of Coulomb force is essential.  

The study of Rydberg state creation has attracted recent interest, with the first direct experimental observation of these excited neutrals presented in \cite{b1}.  The main conclusion was said to provide "strong experimental support" for a rescattering process, grouping it together with HHG and NSMI \cite{b1}.  In contrast, our results using only the standard assumptions from strong field physics (namely the semiclassical tunneling model \cite{b3,b7}) show close quantitative agreement with the same experimental data {\it {without the need to account for rescattering}}.  Moreover, in contradiction to expectation that the inclusion of Coulomb force on the dynamics is essential, we analytically show that Rydberg yield dependence on ellipticity can be closely reproduced while neglecting the influence of the Coulomb force on the trajectory during the entire duration of the long laser pulse.

 Rydberg states can be created when the tunneled electron does not gain sufficient energy from the laser pulse and is subsequently captured by the Coulomb field, creating an excited neutral atom.  
While some progress has been made in analyzing the generation of Rydberg states in linearly polarized light \cite{b20,b21}, the role of polarization, and in particular the dramatic decline of Rydberg states with increasing ellipticity of light remained to be quantifiably explained.  Here, we correct and clarify this dependence of Rydberg yield on ellipticity, deriving a probability distribution which shows excellent agreement with the above mentioned experimental \cite{b1} and numerical data, and which furthermore quantifies the dependence of Rydberg states on laser field intensity.  

We then present our own experimental data showing a hitherto unobserved bifurcation structure in electron energy distribution at the detector.  
This bifurcation in the electron energy from zero coincides with the cut-off for generation of Rydberg states for electrons born in the continuum with zero velocity (corresponding to the peak of the probability distribution).  We show that the value of ellipticity at which this bifurcation occurs can be analytically derived and depends on the interplay between the Coulomb force and the laser field.  

The electric field of a laser propagating in the z-direction is given by:
\begin{equation}
\vec{F}(t) = \frac{-F_0 f(t)}{\sqrt{\epsilon^2 +1}}
\lbrack \cos \left( \omega t + \phi_c \right) \hat{x} + \epsilon sin \left(\omega t + \phi_c \right) \hat{y} \rbrack
\label{eq:field}
\end{equation}
where $\omega$ is the frequency of the laser, $\epsilon$ is the ellipticity (the major axis of polarization is along $\hat{x}$), and $f(t)$ is the slowly-varying pulse envelope: $f_{max} = f(0) =1$.  For many-cycle pulses, we can approximate the carrier-envelope offset (CEO) phase \cite{b17}: $\phi_c \approx 0$.  

To lowest order, the electron dynamics following tunnel ionization are given in atomic units by \cite{b2}: $\partial^2 \vec{r} /\partial t^2 = - \vec{F}(t)$.  This simple equation is based on the assumption within the Strong Field Approximation (SFA) \cite{b2,b4}, that the Coulomb force after ionization is negligible compared to the laser field: $|\vec{F}| \gg Q/r$.
 For the evaluation of Rydberg states, which by definition depend on the Coulomb potential, we will therefore only neglect the Coulomb force up till the end of the laser pulse.  After the laser pulse has passed,  for $t > \tau_p$ (where $\tau_p$ is the pulse duration), we have $f(t) \approx 0$, so that from Eqn. (\ref{eq:field}): $\vec{F}(t) \approx 0$ and therefore the Coulomb force is no longer negligible compared to the laser field, capturing lower energy trajectories into Rydberg states.  
 
For a slowly-varying envelope, where $d f (t)/dt \ll \omega$, the force equation can be integrated to give approximate dynamics for electrons following ionization,
\begin{equation}
x(t) \approx  \frac{F_0 \left(1 - f(t) cos \left(\omega t\right)\right)} {\omega^2 \sqrt{\epsilon^2 + 1} }+ v_{x0} t + x_e
\label{eq:x} 
\end{equation}
\begin{equation}
y(t) \approx  \frac{\epsilon F_0 \left( \omega \triangle t - f(t) sin \left( \omega t \right) 
\right)}{\omega^2 \sqrt{\epsilon^2 + 1}} + v_y \triangle t; \quad z(t) \approx v_z \triangle t
\label{eq:y}
\end{equation}
where $\triangle t = t-t_i$ (with $t_i$ being the instance of ionization), and $x_e$ is the exit point from the tunnel  \cite{b2,b29}.  As an initial condition, we used $v_x\left(t_{i}\right) = 0$ (in accordance with the tunneling model \cite{b7}), which is valid for small $\epsilon$.
The parallel velocity after the laser pulse has passed, $v_{x0}$, is determined by the phase of the field, $\phi_i$, at the instance of ionization: $v_{x0} = -\phi_{i} F_0/\omega \sqrt{1+\epsilon^2}$.  

The transverse velocity at the exit point is given by $\left(v_y, v_z\right)$ and has a Gaussian probability distribution given by the well-known ADK formula \cite{b7}, with standard deviation (SD):  
\begin{equation}
\sigma_\perp = \sqrt{\omega / 2 \gamma}
\label{eq:sd}
\end{equation} 
where $\gamma$ is the Keldysh parameter:  $\gamma=\omega \sqrt{2 I_p}/ F$.

Equation (\ref{eq:x}) shows that an electron that tunnels near the peak, where $v_{x0} \approx 0$, will continue, when averaged over a single cycle, to get farther from the exit point during the duration of the pulse.  
This is because $f(t)$ is a monotonically decreasing function for $t > 0$.  In particular, this means that most Rydberg states, which form predominantly near the peak of the laser field (with formation at the peak corresponding to a dominant quantum number of n=8 for Helium \cite{b1}),  will not rescatter.  

The overall structure of the following derivation for Rydberg yield is:  We show that the electrons end up far from the exit point, $x_e$, after the laser pulse has passed.  This imposes a near
zero final momentum condition.  While the momentum along $x$ is close to zero due to ionization near the peak, the final transverse momentum at the exit point needs to cancel out the drift velocity created by the laser field.  Therefore, the probability of a Rydberg state can be obtained from the probability that the transverse velocity at the exit point (which is given by the ADK formula \cite{b7}) is such that it approximately cancels  the deterministic drift velocity.

From Eqn. (\ref{eq:x}), the position of the electron ionized near the peak after the pulse has passed is:
\begin{equation}
x_f =   \frac{F_0}{\omega^2 \sqrt{\epsilon^2 + 1}} + x_e \sim  \frac{F_0}{\omega^2 }
\label{eq:xf}
\end{equation}
The right-hand side of the equation holds for small $\epsilon$, and in the framework of the strong field infrared laser physics where $F_0/\omega^2  \gg x_e$. 

From Eqn. (\ref{eq:xf}),  it is clear that, in contrast to ionization by a plane wave, an electron ionized near the peak of a laser pulse will end up a significant distance from the parent atom after the laser pulse has passed.  Since the distance from the parent atom at the end of the pulse is large, $x_f \gg x_e$, the velocity has to be small to satisfy the negative energy condition: $-Q/x_f + KE \leq E < 0$, necessary for Rydberg states.  We therefore require that the transverse velocity of the electron after the pulse has passed satisfies the following condition, 
\begin{equation}
\tilde{v}_y = v_y +  v_{drift} \approx 0
\label{eq:vy}
\end{equation}
where $\tilde{v}_y$ was obtained from Eqn. (\ref{eq:y}), and 
\begin{equation}
v_{drift} = \epsilon F_0/\omega \sqrt{\epsilon^2 + 1}
\label{eq:drift}
\end{equation} 
is the drift velocity caused by the vector potential of the laser field \cite{b2,b4}.  

\begin{figure}[t!]
  \includegraphics[width=7cm]{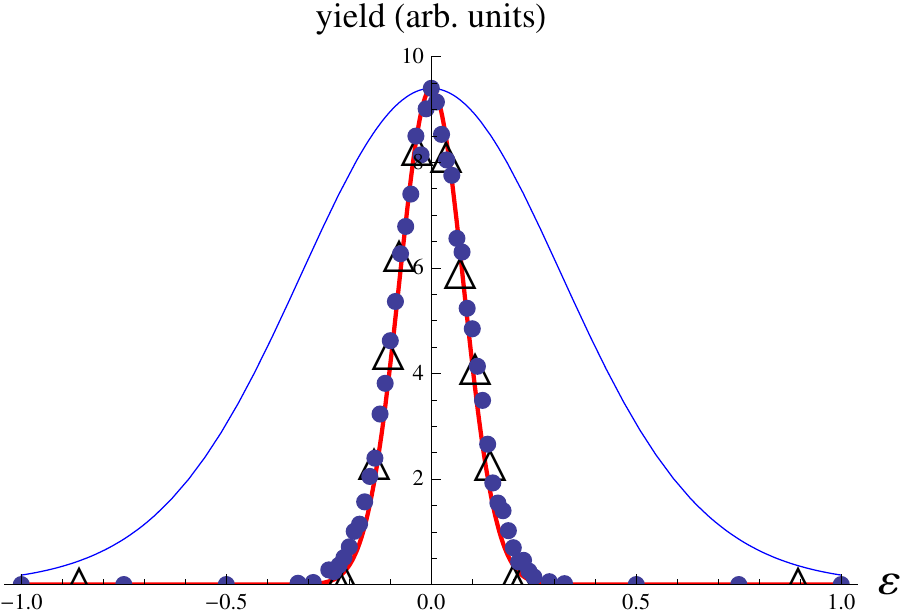}\\
  \caption{Analytic curve (red line), simulations ($\bullet$) and experimental data ($\bigtriangleup$) for the Rydberg yield.  Total ion yield of $He^+$, given by $P_0$, is shown on the same plot (blue line) .  Experimental data is taken from \cite{b1}.  $\tau_p = 30 fs$, $\omega=0.056$ a.u., Intensity $ = 1 PW/cm^2$ }
  \label{Fig2}
\end{figure}

\begin{figure}[t!]
  \includegraphics[width=7cm]{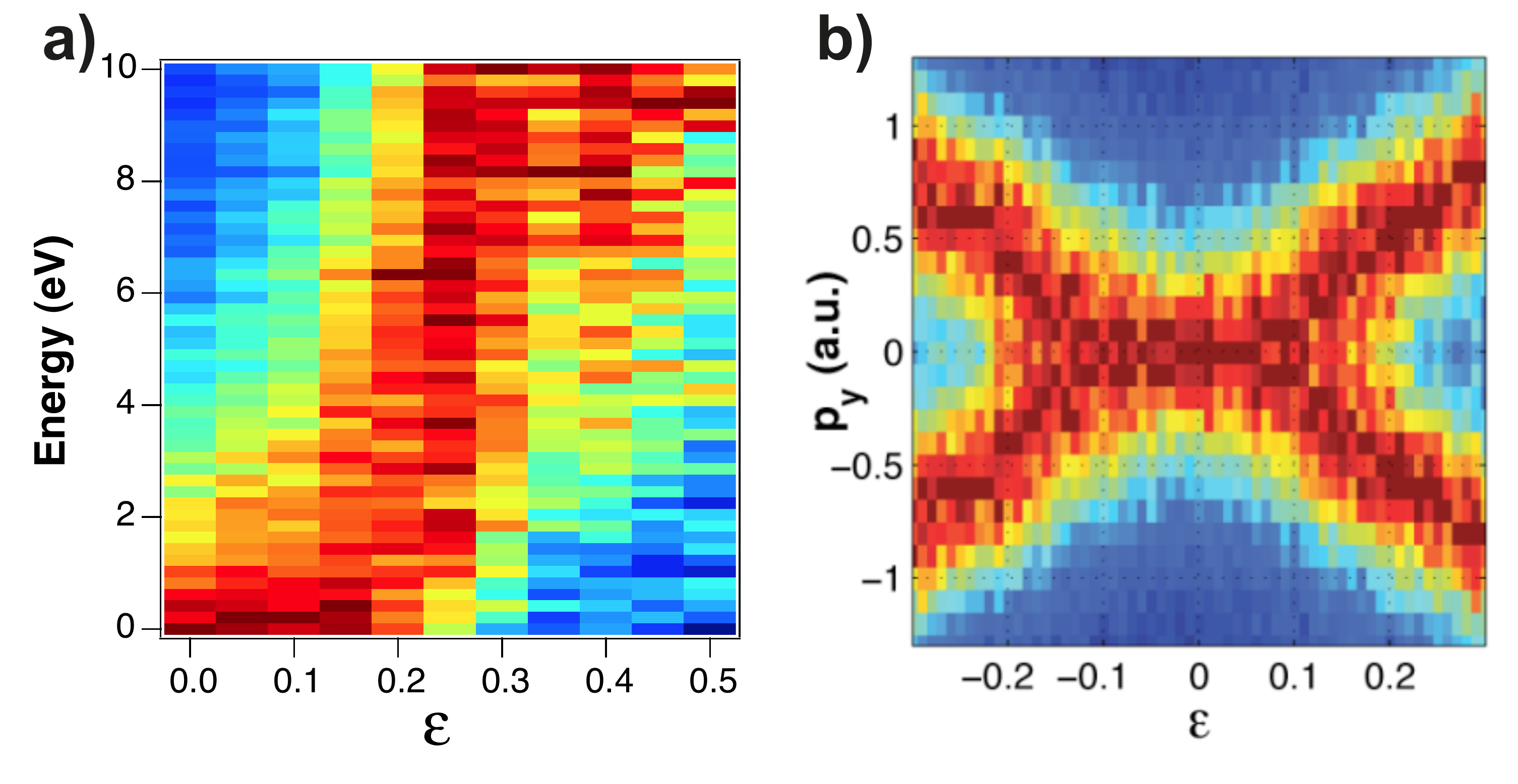}\\
  \caption{a) Energy of electrons observed at the detector as a function of $\epsilon$.  b) Corresponding momentum, $p_y$ along the minor axis of polarization.  $\tau_p=33 fs$, $\lambda_0=788nm$,  Intensity $ = 0.8 PW/cm^2$.}
  \label{Fig3}
\end{figure}

\begin{figure}[t!]
  \includegraphics[width=7cm]{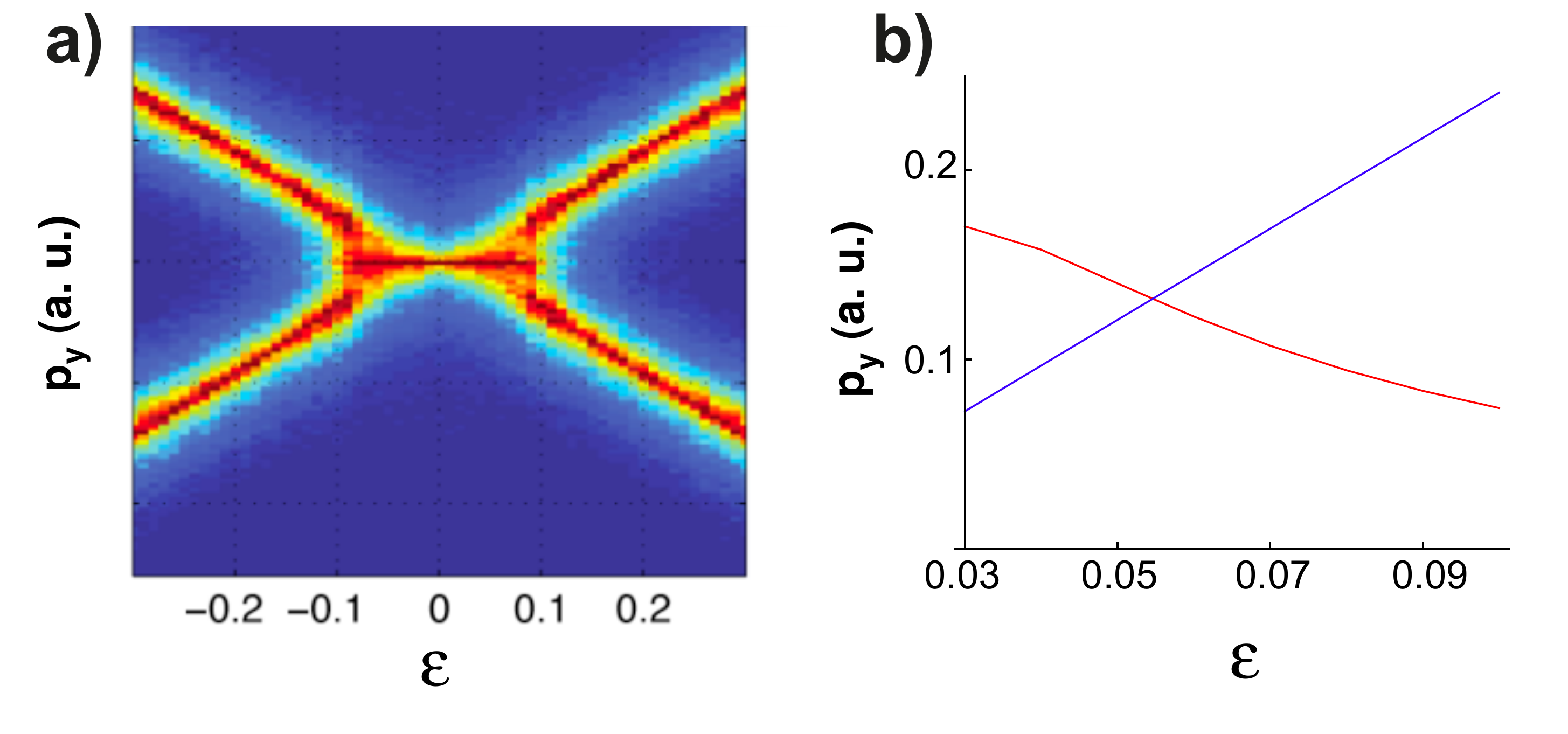}\\
  \caption{(a) Monte Carlo simulations showing a bifurcation of $p_y$ near $|\epsilon| \approx 0.1$, in agreement with experimental results in Fig. \ref{Fig3} (b).  (b) Bifurcation plot.  Red: Coulomb correction to the momentum, $p_{cy}$, given by Eqn. (\ref{eq:corr}).  Blue: drift momentum given by Eqn. (\ref{eq:drift}).  Bifurcation is estimated to occur at the intersection. }
  \label{Fig4}
\end{figure}

From  Eqn. (\ref{eq:sd}), the scaled probability of generating an electron with $v_y \approx -v_{drift}$,
 satisfying Eqn. (\ref{eq:vy}), is $P_R \left(\epsilon\right) \approx  P_0 exp \left(-\epsilon^2 F_0 (2 I_p)^{1/2}/\omega^2 \right)$,
where  $P_0$ is the ionization probability in the tunneling regime along the major axis of polarization \cite{b3}.  In the range of small $\epsilon$, it can be approximated to exponential accuracy as:
$P_0 = exp \left(-2 (2I_p)^{3/2} \sqrt{1+\epsilon^2}/3 F_0 \right)$ \cite{b30}.  The exponent of $P_0$ can be further Taylor expanded to obtain,
\begin{equation}
P_R \left(\epsilon\right) \approx e^{\frac{-2\left(2 I_p\right)^{3/2}}{3 F_0}} e^{-\epsilon^2/2 \sigma_{\epsilon}^2}
\label{eq:pR2}
\end{equation}
where $\sigma_{\epsilon}$ is the standard deviation of the Gaussian probability distribution for Rydberg state yield as a function of ellipticity of light:
\begin{equation}
\sigma_\epsilon = \sqrt{\frac{3}{3+\gamma^2}} \frac{\omega}{\sqrt{2 F_0} \left(2 I_p \right)^{1/4}}
\label{eq:SDep}
\end{equation}
 In deriving Eqn. (\ref{eq:pR2}) a Taylor expansion was used, retaining only terms up to the order of $\epsilon^2$, since $\epsilon^2 \ll 1$ (the Rydberg state generation is only found at small values of $\epsilon$). 
In the tunneling regime where $\gamma \ll 1$, Eqn. (\ref{eq:SDep}) can be approximated as:  $\sigma_{\epsilon} = \omega/\sqrt{2 F_0} (2 I_p)^{1/4}$. 

The SD given in Eqn. (\ref{eq:SDep}) increases with decreasing $F_0$, predicting a slower decline of Rydberg states with $\epsilon$, as laser intensity goes down.  This is in contrast to the total ionization yield, given by $P_0$, where smaller laser intensities lead to a faster decline with ellipticity.
These two factors lead to an increase of Rydberg trajectories relative to the total ionization yield, explaining and quantifying the findings in \cite{b1}, where numerics show ``an increasing percentage of bound trajectories ... with decreasing laser intensity."   On the other hand, for linearly polarized light, where $\epsilon =0$, Eqn. (\ref{eq:pR2}) predicts an increase in Rydberg states with increasing intensity, also in agreement with the experimental and numerical findings in \cite{b1}.

Monte Carlo simulations were performed whereby an ensemble of $10^5$ ionization events with a randomly chosen CEO phase were generated for each value of $\epsilon$ over the interval [-1,1].  The ionization rate for each event is given by Tong et al. \cite{b18}, assuming instantaneous tunneling delay time \cite{b10}.  The equation of motion is solved with the initial condition that the electron starts its trajectory outside the potential barrier, at the exit point, $x_e$,  with an initial longitudinal momentum of zero and a transverse momentum distribution given by Eqn. (\ref{eq:sd}).

A comparison between the experimental data (taken from \cite{b1}), simulations and analytically obtained curve given by Eqn. (\ref{eq:pR2}) is shown in Fig. \ref{Fig2}.  The intensity and laser frequency used in the figure were chosen for the purposes of comparison with the experiment in \cite{b1}.  While the total ion yield varies slowly with increasing ellipticity, the yield of Rydberg states drastically decreases to essentially zero for $\epsilon > 0.3$.  This holds true over a wide span of intensities ($0.35 - 3 PW/cm^2$), where analysis (see Eqn. (\ref{eq:pR2})) and numerics show the disappearance of Rydberg states in the $0.2 < \epsilon < 0.3$ range, depending on field strength.  
This dramatic decrease with ellipticity is also typical of a rescattering process as measured for HHG and NSMI \cite{b1, b5} .  

{\it{Experiment}}:  
Our experiment provides indirect information on Rydberg state generation by measuring the yield of $E \approx 0$ states at the detector as a function of $\epsilon$.  $E \approx 0$ states can be considered as a cut-off for Rydberg states  since the condition defining the Rydberg state (with the corresponding quantum number, $n$) is given by:  $E = -1/2n^2 < 0$. 

The experimental setup was as follows:  
a laser pulse of duration $\tau_p = 33 fs$, central wavelength $\lambda_0 = 788nm$,  and peak intensity of $0.8 PW/cm^2$ 
(CEO phase \cite{b17} was not stabilized)
 was produced by a Ti:Sapphire laser system focused onto helium atoms in a cold gas jet, with the gas jet density adjusted such that on average much less then one ionization occurs per laser shot.  A COLTRIMS setup \cite{b10} measures the ion momentum, which is the negative of the electron momentum due to momentum conservation.  The momentum resolution is 0.1 a.u. in time-of-flight direction and 0.9 a.u. in gas jet direction, mainly determined by thermal spread.  A broadband quarter-wave plate is used to alter the ellipticity of the laser pulses.  In the final analysis, the ellipticity and the angular orientation of the polarization ellipse are calculated from the angle of the quarter-wave plate, respecting its wavelength dependence \cite{b15}.  The knowledge of the ellipticity and the angle of the polarization ellipse for each detected ion allow generating ellipticity-resolved spectra with a high resolution.
 
The experimental results are presented in Fig. \ref{Fig3}.   Figure \ref{Fig3} (a) shows that $E \approx 0$ states decline dramatically and virtually disappear for $\epsilon > 0.3$, in agreement with theory and numerics, shown in Fig. \ref{Fig2}, which also predict the disappearance of Rydberg states close to $\epsilon \approx 0.3$.  

The experimental data in Fig. \ref{Fig3} (b), and the corresponding Monte Carlo simulation in Fig. \ref{Fig4} (a), show an interesting bifurcation behavior where the center of the $p_y$ distribution, where $p_y$ is the momenta along the minor axis of polarization, splits from $p_y \approx 0$ as $\epsilon$ is increased.  (By bifurcation, we mean a qualitative change in behavior as one of the parameters, in this case $\epsilon$, is increased.)  This bifurcation can not be explained within the SFA approximation, which neglects Coulomb force after ionization, predicting a close to linear increase in $|p_y|$ as $|\epsilon|$ increases  \cite{b31}, which would result in  immediate splitting as $\epsilon$ is increased from zero.
To explain the bifurcation behavior shown in Fig. \ref{Fig3} (b), we therefore have to take account of the Coulomb correction to the electron dynamics after ionization.  Using perturbative approach \cite{b28}, the total Coulomb correction to the momentum is estimated by assuming that to lowest order the electron follows a trajectory given by Eqns. (\ref{eq:x}) and (\ref{eq:y}).

The most probable trajectory corresponds to ionization at the peak with $v=0$, and is given by $\vec{r}_0 (t) =x_0(t) \hat{x} + y_0(t) \hat{y}$, where $x_0(t)$ and $y_0(t)$ are given by Eqns. (\ref{eq:x}) and (\ref{eq:y}) with $v_x=v_y=0$.  The total Coulomb correction  to the center of the distribution is \cite{b28},
\begin{equation}
\vec{p}_c = - \int_{0}^{\tau_p} dt \left(\frac{x_0(t) \hat{x} + y_0(t) \hat{y} }{r_0^3(t)}\right) =  p_{cx} \hat{x} + p_{cy} \hat{y}
\label{eq:corr}
\end{equation}
where the integration is from the time of ionization at the peak, where $t_i=0$, till the end of the pulse at $t \approx \tau_p$.

The total momenta along $y$, as observed at the detector, is: $p_y =  v_{drift} + p_{cy}$.  
A non-zero drift of the center of the distribution occurs for $|p_y| > 0$, with the corresponding bifurcation around: $v_{drift} = - p_{cy}$.  Using Eqns. (\ref{eq:y}),  (\ref{eq:drift}) and (\ref{eq:corr}), 
\begin{equation}
\frac{1}{\omega} \int_0^{T/2}  dt \frac{\omega t - f(t) sin(\omega t)}{\left(x_0^2(t, \epsilon_b)+y_0^2(t, \epsilon_b)\right)^{3/2}} =1
\label{eq:offset}
\end{equation}
where, as before, $x_0(t,\epsilon_b)$ and $y_0(t,\epsilon_b)$ are given by Eqns. (\ref{eq:x}) and (\ref{eq:y}) with $v_x$, $v_y$, and $t_i$ set to zero.  

Equation (\ref{eq:offset}) can be solved numerically for the bifurcation value, $\epsilon_b$.  Using the same parameters as for the experimental data shown in Fig. \ref{Fig3}, we get a bifurcation value of $\epsilon_b \approx 0.06$.  
This estimate is in approximate agreement with experimental data,  (Fig. \ref{Fig3} (b)) and with
numerics (Fig. \ref{Fig4} (a)), which both show a bifurcation near $\epsilon \approx 0.1$.  
As shown in Fig. 3 (b), the bifurcation point corresponds to the value of $\epsilon$ where the Coulomb force becomes too weak to offset the velocity drift along the minor axis of polarization created by the laser field.   An under-estimate of $\epsilon_b$ is expected when using a perturbative approach since an electron that follows the unperturbed path, $\vec{r}_0$, feels no Coulomb force and therefore gets away from the atom faster than the actual trajectory, thereby incurring less of a Coulomb correction.  

In conclusion, we analyzed the dependence of Rydberg states on laser ellipticity, deriving a Gaussian probability distribution for the yield of neutral Rydberg atoms as a function of $\epsilon$; this puts recent experimental and numerical results into a theoretical framework.  In particular, our work suggests that rescattering does not play a significant role in the creation of excited neutrals, in contradiction to the mechanism proposed in \cite{b1}.

Our analysis predicts the disappearance of Rydberg states for polarizations with $\epsilon >0.3$ (when the electron yield from tunnel ionization is still quite substantial).  The standard deviation of the derived probability distribution, given by Eqn. (\ref{eq:SDep}),  is found  to decrease with increasing laser intensity, in agreement with the observed decline of Rydberg states at higher laser intensities.  

The experimental data presented here confirms the disappearance of Rydberg states for $\epsilon  > 0.3$ by measuring zero energy states at the detector.  These states can be considered as a cut-off for the creation of bound trajectories since they correspond to asymptotically large values of $n$ (the quantum number of the Rydberg state).  In addition, our experiment shows a bifurcation from zero in the final momentum, $p_y$, as ellipticity is increased.  This bifurcation was explained by a perturbative inclusion of a Coulomb correction. 

We gratefully acknowledge Dr. Eichmann for providing us with experimental data displayed in Fig.1 (see \cite{b1}), the NCCR Quantum Photonics (NCCR QP), ETH Research Grant ETH-03 09-2, and the EU FP7 grant.


\begin{thebibliography}{99}

\bibitem{b1}	T. Nubbemeyer, K. Gorling, A. Saenz, U. Eichmann, and W. Sandner, Phys. Rev. Lett. {\bf 101}, 233001 (2008).
\bibitem{b3}	L.V. Keldysh, JETP {\bf 20}, 1307 (1965).
\bibitem{b2}	P.B. Corkum, Phys. Rev. Lett. {\bf 71}, 1994 (1993).
\bibitem{b34} M. Lewenstein, {\it{et al}}, Phys. Rev. A  {\bf 49}, 2117 (1994).
\bibitem{b35} A. Zair, {\it{et al}}, Phys. Rev. Lett. {\bf 100}, 143902 (2008).
\bibitem{b5}	P. Dietrich, N.H. Burnett, M. Ivanov, and P.B. Corkum, Phys. Rev. A {\bf 50}, R3585 (1994).
\bibitem{b36} A.N. Pfeiffer, C. Cirelli, M. Smolarski, R. Dorner, U. Keller, Nature Physics {\bf 7}, 428 (2011).
\bibitem{b37} A.N. Pfeiffer, {\it{et al}}, New J. Phys. 13, 093008 (2011).
\bibitem{b23} T. Brabec, M.Y. Ivanov, and P.B. Corkum, Phys. Rev. A {\bf 54}, R2551 (1996).
\bibitem{b25}  C.I. Blaga  {\it{et al}}, Nat. Phys. {\bf 5}, 335 (2009).
\bibitem{b28} S.P. Goreslavski, G.G. Paulus, S.V. Popruzhenko, and N.I. Shvetsov-Shilovski, Phys. Rev. Lett. {\bf 93}, 233002 (2004).
\bibitem{b29}	A.N. Pfeiffer et al., Nature Physics 8, 76-80 (2012).
\bibitem{b21}    G.L. Yudin and M.Y. Ivanov, Phys. Rev. A {\bf 63}, 033404 (2001).
\bibitem{b7}	M.V. Ammosov, N.B. Delone, and V.P. Krainov, Sov. Phys. JETP {\bf 64}, 199 (1986).
\bibitem{b20}	N.I. Shvetsov-Shilovski {\it{et al}}., Laser Phys. {\bf 19}, 1550  (2009).
\bibitem{b17}	H. R. Telle et al., Appl. Phys. B {\bf 69}, 327 (1999).
\bibitem{b4}	M.Y. Ivanov, M. Spanner, and O. Smirnova, J. of Modern Optics {\bf 52:2}, 165 (2005).
\bibitem{b30}  A.M. Perelomov, V.S. Popov, and M.V. Terent'ev, JETP 50, 1393 (1966).
\bibitem{b18}	X. M. Tong and C. D. Lin, J. Phys. B {\bf 38}, 2593 (2005).
\bibitem{b10}	P. Eckle et al., Science {\bf 322}, 1525 (2008).
\bibitem{b15}	M. Smolarski et al., Opt. Express {\bf 18}, 17640 (2010).
\bibitem{b31}  V.S. Popov, Physics-Uspekhi 47 (9) 855-885 (2004).


\end{thebibliography}
\end{document}